\documentclass[10pt,aps,pre,reprint,nofootinbib]{revtex4-1}

\usepackage{amssymb,amsmath}
\usepackage{graphicx}
\usepackage{color}
\usepackage{hyperref}
\usepackage{soul}
\hypersetup{
    pdfauthor={AIB},     
    colorlinks=true,       
    linkcolor=blue,          
    citecolor=red,        
}
\graphicspath{{figures/}}

\def\F1{F_1}

\def\md{\mathrm d}
\def\kT{k_{\rm B}T}
\def\xT{x_{\textrm{trap}}}
\def\half{\tfrac{1}{2}}
\def\Edim{E'}
\def\kdim{k'}
\def\xdim{x'}
\def\tdim{t'}
\def\tC{t_{\rm c}}
\def\DeltaEdim{\Delta E'}
\def\Escale{H}
\def\lscale{\ell}
\def\tscale{\tau}
\def\ALR{A_{\textrm{LR}}}
\def\ARoom{A_{\textrm{room}}}
\def\ALimit{A_{\textrm{limit}}}

\newcommand{\ltsim}{\protect\raisebox{-0.5ex}{$\:\stackrel{\textstyle <}{\sim}\:$}}
\newcommand{\gtsim}{\protect\raisebox{-0.5ex}{$\:\stackrel{\textstyle >}{\sim}\:$}}

\begin{document}
\title{Effective dissipation: breaking time-reversal symmetry in driven microscopic energy transmission}
\author{Aidan I. Brown}
\email{aidanb@sfu.ca}
\author{David A. Sivak}
\email{dsivak@sfu.ca}
\affiliation{Department of Physics, Simon Fraser University, Burnaby, BC, V5A1S6 Canada}
\date{\today}

\begin{abstract} 
At molecular scales, fluctuations play a significant role and prevent biomolecular processes from always proceeding in a preferred direction, raising the question of how limited amounts of free energy can be dissipated to obtain directed progress. We examine the system and process characteristics that efficiently break time-reversal symmetry at fixed energy loss; in particular for a simple model of a molecular machine, an intermediate energy barrier produces unusually high asymmetry for a given dissipation. We relate the symmetry-breaking factors found in this model to recent observations of biomolecular machines.
\end{abstract}
\maketitle

\section{Introduction}Biomolecular processes are generally out of equilibrium~\cite{battle16,kolomeisky07,andrieux06}, hindering our quantitative understanding of their operation.
Molecular machines typically operate far from equilibrium in converting between different forms of energy to perform various cellular tasks. 
Transport motors, such as kinesin~\cite{block07,carter05,clancy11}, use ATP hydrolysis~\cite{schnitzer97} to bias motion in a particular direction~\cite{roldan14}, but random fluctuations cause microscopic biomolecular machines to sometimes operate backwards~\cite{carter05}.

A process with zero dissipation must be reversible.
Thus net progress in a preferred direction 
requires some dissipation~\cite{vandenbroeck10,roldan14,machta15,barato15}, but the Second Law is silent as to more quantitative details. For a simple model of a driven process, we investigate how much irreversible progress can be achieved for a given amount of free energy dissipation and identify process properties that produce asymmetry approaching the fundamental physical limits.

The breaking of time symmetry (leading to ``time's arrow'')~\cite{feng08} is naturally expressed 
by the Jensen-Shannon divergence between the 
trajectory ensembles for forward and time-reversed driven processes.  Information-theoretic inequalities set upper limits on the time asymmetry for a given dissipation, increasing monotonically with dissipation~\cite{feng08}; essentially, achieving a given difference between forward and reverse dynamics requires paying a certain minimum cost (in dissipation), averaged over all realizations.
Such broken time-reversal symmetry is related to spatially anisotropic biomolecular motion and functionally asymmetric machine operation (\emph{e.g.}, synthesizing not hydrolyzing ATP).
To the extent that a given biomolecular component was sculpted by natural selection to achieve directed progress yet avoid unnecessarily wasting energy, it may, 
subject to physical limits, 
achieve abnormally high time asymmetry given a particular `dissipation budget'~\cite{shoval12,tawfik14,lan12}.

Existing empirical explorations (experimental RNA hairpin unfolding/refolding~\cite{collin05} and molecular dynamics simulations unfolding/refolding alanine decapeptide~\cite{procacci10}) give time asymmetry-dissipation tradeoffs similar to that of a generic system obeying linear response theory~\cite{marconi08}. At low (0 -- 1 $\kT$) and high ($\gtsim 8$ $\kT$) dissipation the linear response system's time asymmetry nears the upper limit (Fig.~\ref{fig:asymmDiss}), but at intermediate dissipations the linear response time asymmetry lies well below the theoretical maximum, leaving room for improvement.

While the Second Law and information theory set absolute limits on time asymmetry for a given dissipation, we currently lack understanding of the factors that increase or decrease the time asymmetry within these limits. 
This paper addresses, for the first time, characteristics of a system or process that increase time asymmetry at a given dissipation. We investigate this issue in a model nonequilibrium process, a harmonic trap translating at a constant velocity and thereby dragging a diffusing particle over a step barrier. We find the largest time asymmetry values at a given dissipation occur for intermediate step height and a close initial proximity to the step.

To understand the determinants of enhanced time asymmetry at a given dissipation, we examine protocols where the instantaneous probability distribution differs substantially from the corresponding equilibrium distribution, true for sufficiently rapid protocols. In particular, intermediate step heights lead to far-from-equilibrium distributions when ascending the step, but near-equilibrium distributions when descending the step. This produces significant time asymmetry.
A smaller step height is insufficient to drive the system far from equilibrium in either direction,
while a larger step height 
keeps the equilibrium distribution always below the step, preventing any significant nonequilibrium lag. 

\section{Methods}The Crooks fluctuation theorem~\cite{crooks99} implies that the work 
associated with a given trajectory captures all information about the relative probabilities of appearing in the forward or reverse trajectory distributions.
The time asymmetry $A$ can be precisely estimated from the empirical forward and reverse work distributions~\cite{feng08},
\begin{equation}
\label{eq:asymmetry}
\begin{split}
A[\Lambda] \equiv \half\left\langle\ln\frac{2}{1 + \exp\left(-\beta W[x|\Lambda] + \beta\Delta F\right)}\right\rangle_{\Lambda}\\
+ \half\left\langle\ln\frac{2}{1 + \exp\left(-\beta W[\tilde{x}|\tilde{\Lambda}] - \beta\Delta F\right)}\right\rangle_{\tilde{\Lambda}}.
\end{split}
\end{equation}
$\Lambda$ labels a protocol, the time course of a controllable parameter $\lambda(t)$ over $t\in [0,\Delta t]$, for duration $\Delta t$. $\tilde{\Lambda}$ is the time-reversal of $\Lambda$. $W[x|\Lambda]$ is the work done during system trajectory $x$ subject to protocol $\Lambda$. $\beta\equiv(\kT)^{-1}$ and $\Delta F$ is the free energy change over the forward protocol $\Lambda$. If one observes a single trajectory resulting from either a particular nonequilibrium process or its time reversal (with 50\% prior probability of either), $A$ quantifies the expected information gain, in nats ($1/\ln 2$ bits), about whether the trajectory was produced from a forward or reverse protocol~\cite{crooks11}.
For identical trajectory distributions, any observation provides no information on the direction of time, defining a minimum $A=0$. Completely distinct trajectory distributions guarantee that any observed trajectory can be definitively assigned to the forward or reverse process, defining a maximum $A=\ln 2$ nats, corresponding to one bit of information.

Here we define dissipation as the average dissipated work for a uniform mix of forward and reverse protocols, and hence (since they have opposite free energy changes) the average work~\cite{feng08}
\begin{equation}
\label{eq:dissipation}
h[\Lambda] \equiv \half\beta\left\langle W[x|\Lambda]\right\rangle_{\Lambda} + \half\beta\left\langle W[\tilde{x}|\tilde{\Lambda}]\right\rangle_{\tilde{\Lambda}} \ .
\end{equation}
For a given dissipation $h$, the time asymmetry $A$ cannot exceed certain limits~\cite{feng08, taneja05} (the grey forbidden region  in Fig.~\ref{fig:asymmDiss}): $A\leq h/4$ and $A\leq\ln[2/(1+e^{-h})]$. The linear response $A$ vs.\ $h$ relationship~\cite{feng08} (solid black curve in Fig.~\ref{fig:asymmDiss}) is determined from Gaussian work distributions with mean dissipation $\langle W\rangle - \Delta F = \half \sigma_W^2$, for work variance $\sigma_W^2$ \cite{feng08,speck04}.

We investigate these issues in a model system, an overdamped particle with diffusivity $D$ on a potential landscape $E(x,\xT(t)) = E_{\textrm{trap}}(x,t) + E_{\textrm{step}}(x)$ composed of two components: a spring represented by a quadratic potential, $E_{\textrm{trap}}(x,t) = \half k\left(x - \xT(t)\right)^2$, with a time-dependent minimum $\lambda = \xT(t)$ as the control parameter; and a step potential at $x=0$, $E_{\textrm{step}}(x) = \Theta(x)\Delta E$, for Heaviside function $\Theta(x)$. The step represents an energetically unfavorable transition or energy storage stage~\cite{derenyi99}. $E$, $k$, $x$, and $\Delta E$ are dimensionless quantities --- notably, $E$ and $\Delta E$ are in units of $\kT$ (non-dimensionalization details in Appendix \ref{sec:nondim}).

The trap translates at constant velocity $u=\Delta x/\Delta t$ from $\xT^{\rm i} = -\Delta x/2$ to $\xT^{\rm f} = \Delta x/2$ for the forward protocol, and in the opposite direction for the reverse protocol. The particle begins in equilibrium and diffuses as the trap translates. 
Further numerical simulation details are in Appendix \ref{sec:numdet}.

For a given spring constant $k$, step height $\Delta E$, distance $\Delta x$, and velocity $u$, work distributions are accumulated from many repetitions of the forward and reverse protocols, from which a single time asymmetry (Eq.~\ref{eq:asymmetry}) and dissipation (Eq.~\ref{eq:dissipation}) are calculated. Repeating this for varying velocities produces a parametric curve of time asymmetry vs.\ dissipation for a given spring constant, step height, and distance.

\section{Results}
\begin{figure}[tbp] 
	\centering
	\hspace{-0.3in}
	\begin{tabular}{c}
		\hspace{0.200in}\includegraphics[width=3.375in]{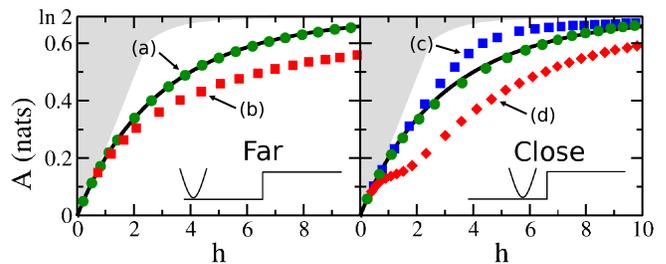}\\
	\end{tabular}
	\caption{\label{fig:asymmDiss} 
{\bf Time asymmetry $A$ (Eq.~\ref{eq:asymmetry}) vs.\ dissipation $h$ (Eq.~\ref{eq:dissipation}).} Grey region shows unfeasible time asymmetries for a given dissipation. Solid black curve shows linear response. Left: `Far' protocols, where a trap with spring constant $k=10$ begins $\Delta x/2=10$ away from a step of height $\Delta E = 4$ (green circles) or $9$ (red squares). Right: `Close' protocols, where a trap with $k=10$ begins $\Delta x/2 = 1.5$ away from a step of height $\Delta E = 4$ (green circles), $9$ (blue squares), or $14$ (red diamonds). Arrows indicate work distributions shown in Fig.~\ref{fig:workDists}, with corresponding letter labels. Standard errors, estimated over 10 samples of $10^4$ runs, are smaller than the data points.}
\end{figure}

Fig.~\ref{fig:asymmDiss} shows how different step heights and initial trap positions $\xT$ lead to time asymmetries on, below, and above linear response at a given dissipation. Significant regions of parameter space give results indistinguishable from linear response, and thus the linear response time asymmetry provides a natural baseline with which to compare the time asymmetry of a particular process.

\begin{figure}[tbp] 
	\centering
	\hspace{-0.3in}
	\begin{tabular}{c}
		\hspace{0.10in}\includegraphics[width=3.375in]{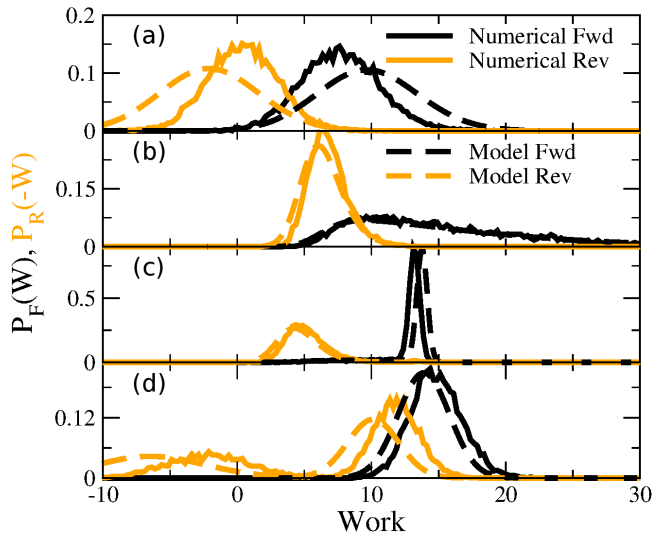} \\
	\end{tabular}
	\caption{\label{fig:workDists}  {\bf Work distributions.} 
Numerical simulations (solid) and semi-analytic calculations (dashed) of forward (black) and reverse (orange) protocols, $P_{\textrm{F}}$ and $P_{\textrm{R}}$, respectively. (a) -- (d) correspond to indicated points in Fig.~\ref{fig:asymmDiss}.
(a) and (b) show far protocols ($\Delta x/2 = 10$). 
(a) Spring constant $k = 10$, step height $\Delta E = 4$, and protocol velocity $u = 1.44\times10^{-1}$.
(b) $k = 10$, $\Delta E = 9$, and $u = 1.00\times10^{-2}$.
(c) and (d) show close protocols ($\Delta x/2 = 1.5$). (c) $k = 10$, $\Delta E = 9$, and $u = 5.04\times10^{-2}$. (d) $k = 10$, $\Delta E = 14$, and $u = 1.33$. Distributions are over $10^4$ runs.
}
\end{figure}

The corresponding work distributions (Fig.~\ref{fig:workDists}) show that a wider original peak and the emergence of a high-work peak can produce decreased or increased time asymmetry, respectively.

The left panel of Fig.~\ref{fig:asymmDiss} shows time asymmetry vs. dissipation for `far' protocols, where the trap begins $\Delta x/2 = 10$ away from the step with spring constant $k = 10$. 
Here the initial equilibrium distribution is unaffected by the potential step. The green circles, simulations with a step height $\Delta E = 4$, follow the solid black linear response curve. For point (a), the forward and reverse work distributions appear symmetric with similar variances, consistent with the Gaussian work distributions expected from linear response behavior (Fig.~\ref{fig:workDists}a).

The red squares, for simulations with step height $\Delta E = 9$, follow the linear response curve at low velocities (corresponding to lower dissipation), but lie significantly below the linear response prediction for sufficiently fast velocities. For point (b) the forward and reverse work distributions are quite distinct (Fig.~\ref{fig:workDists}b): the forward distribution is significantly wider than the reverse, extending to large positive work values. This stretched forward work distribution, along with a non-stretched reverse work distribution, leads to lower time asymmetry at a given dissipation than for linear response.

The right panel of Fig.~\ref{fig:asymmDiss} shows results for `close' protocols, where the trap (with spring constant $k = 10$) begins $\Delta x/2 = 1.5$ away from the step, and hence initial equilibrium distributions are influenced by the potential step. The green circles, simulations with step height $\Delta E = 4$, reproduce the black linear response curve. The blue squares, with step height $\Delta E = 9$, show intermediate velocities with increased time asymmetries. Point (c) has a forward work distribution (Fig.~\ref{fig:workDists}c) 
with a dominant high-work peak 
having little overlap with the reverse low-work peak; hence these distributions are more distinguishable, increasing the time asymmetry at a given dissipation.

The red diamonds, simulations with step height $\Delta E = 14$, fall below linear response at medium-to-high velocities. At point (d) the forward and reverse work distribution (Fig.~\ref{fig:workDists}d) both have peaks at high work values. The close overlap of the
high-work 
peaks produces lower time asymmetry at a given dissipation than for linear response.

\begin{figure}[tbp] 
	\centering
	\hspace{-0.3in}
    \begin{tabular}{c}
    \hspace{0.100in}\includegraphics[width=3.375in]{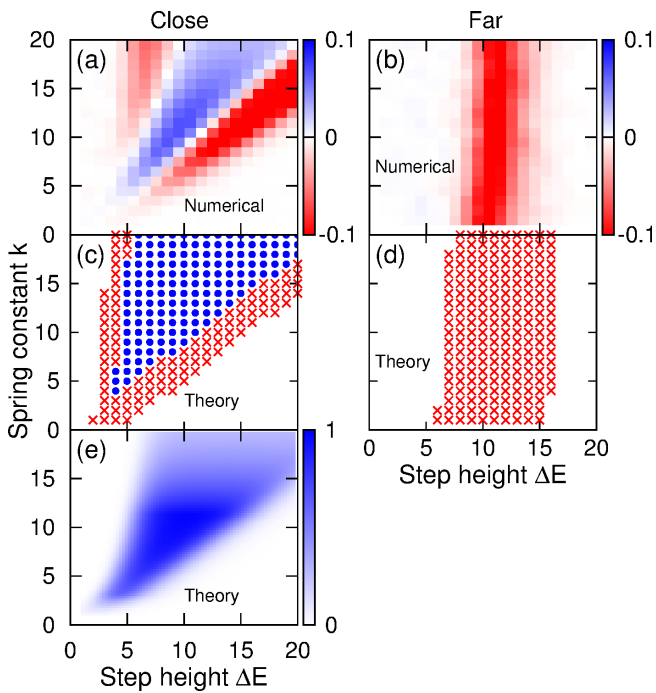}\\
    \end{tabular}
	\caption{\label{fig:heatPhase} {\bf Deviation from linear response.} Varying step height $\Delta E$ and spring constant $k$. (a) and (b) are numerical results for the difference between the resulting time asymmetry $A$ and the linear response value at the same dissipation $\ALR$, $A - \ALR$. (a) is for $\Delta x/2=1.5$ and $u = 1.76\times10^{-1}$, (b) is for $\Delta x/2=10$ and $u = 1.18\times10^{-2}$. (c) and (d) are phase diagrams, corresponding to the same parameters as (a) and (b), respectively, predicted by the criteria described in the text. Red x's represent below linear response, blue circles above linear response, and no symbols on linear response. (e) is the quantity $P_{\textrm{down step}}(1-P_0)\ARoom$ from \ref{sec:maximizing}, which is an analytical prediction of amount above linear response, using the same parameters as (a) and (c).}
\end{figure}

We examine in more detail
the widening of the main peak (e.g. for the forward work distribution in Fig.~\ref{fig:workDists}b) and the origin of the high-work peaks (e.g. the forward distribution in Fig.~\ref{fig:workDists}c and both distributions in Fig.~\ref{fig:workDists}d). 
In Appendix \ref{sec:workdist} we derive semi-analytic work distributions with no free parameters (Fig.~\ref{fig:workDists}, dashed lines) that primarily consider the work done as the trap moves above the step while the particle remains below the step. The qualitative match to the numerical simulations (Fig.~\ref{fig:workDists}, solid lines) suggests this lag is the salient feature producing the variation among work distributions.
Related analysis (Appendix \ref{sec:criteria}) elucidates inequalities that govern when time asymmetry departs significantly from that of linear response (Fig.~\ref{fig:heatPhase}c,d). 

For an overdamped particle in a trap translating at constant velocity on a flat landscape, at steady state the nonequilibrium position distribution lags the equilibrium distribution by a constant distance~\cite{mazonka99}. This system has Gaussian work distributions that obey linear response at any velocity. Alternatively, if the forward and reverse work distributions do not overlap, dissipation tends to be sufficiently large such that the high time asymmetry is indistinguishable from linear response. 
Time asymmetry differs from that of linear response when the nonequilibrium distribution deviates from equilibrium in a different manner, producing non-Gaussian work distributions with some overlap.

During the forward protocol, for sufficiently large step height $\Delta E \gtsim \half\ln \frac{32D^2k}{9 u^2}$, the particle lags and doesn't immediately follow the trap over the step. For large $\Delta x$, this produces a stretched forward work distribution, and hence a time asymmetry below linear response.
For step height $\Delta E \gtsim \ln(Dk\Delta t)$ and intermediate $\Delta x$, the particle remains below the step for the entire forward protocol while the equilibrium distribution shifts past the step. 
Satisfying both inequalities above, along with intermediate $\Delta x$, results in a significant high-work peak, producing a time asymmetry above linear response.

For step height $\Delta E \gtsim \frac{k}{2}\frac{(\Delta x)^2}{2}$, the reverse protocol's initial equilibrium includes significant probability below the step. This produces a significant reverse high-work peak, which combined with an existing forward high-work peak produces a time asymmetry below linear response.
For step height $\Delta E \gtsim \frac{k}{2}\frac{(\Delta x)^2}{2} + \ln(100)$, nearly all the initial equilibrium distribution for the reverse protocol begins below the step, so virtually all realizations in both directions fall in the high work peak, resulting in linear response behavior.

For relatively small $\Delta x$, these inequalities can be fulfilled in the order they are listed above, moving left to right with increasing $\Delta E$ in Fig.~\ref{fig:heatPhase}c. For large $\Delta x$, only the first inequality is fulfilled (Fig.~\ref{fig:heatPhase}d); at even higher $\Delta E$ the forward and reverse work distributions do not overlap at all.

Fig.~\ref{fig:heatPhase} shows time asymmetry phase diagrams for systematic variation of $k$ and $\Delta E$, either (a,b) directly calculated via the previously described numerical simulations, (c,d) predicted via the above inequalities, or (e) predicted via further analysis (Appendix \ref{sec:maximizing}). The qualitative match between the numerical results and our predictions in Fig.~\ref{fig:heatPhase} confirms that the time 
the particle remains below the step can explain the time asymmetry.

The analysis presented in Figs.~\ref{fig:heatPhase}c, d, and e 
predicts that an intermediate range of $\Delta E$ will produce significantly higher time asymmetry than linear response. 
These bounds can also correspond to an intermediate range of protocol distances, with $\sqrt{8\Delta E/k} \ltsim \Delta x \ltsim \exp(\Delta E)/(Dku)$. 

In this system, time asymmetry above linear response requires (1) a significant forward high-work peak, (2) a significant reverse low-work peak, and (3) linear response time asymmetry significantly below the upper limit. Combining these three criteria (Fig.~\ref{fig:heatPhase}e, details in Appendix \ref{sec:maximizing}) produces a maximal time asymmetry gap at roughly the same location as seen numerically (Fig.~\ref{fig:heatPhase}a). 

\section{Discussion}
To explore how small amounts of free energy dissipation can be used to generate directed progress of molecular machines, we investigated time asymmetry as a function of dissipation for a model system: an overdamped particle diffusing in a quadratic trap moved over a potential step. 
For step heights significantly larger than $\kT$, the time asymmetry for a given dissipation can depart from that of linear response, either above or below.
 
Exceeding linear response time asymmetry requires intermediate step heights, which manifest lag when ascending the step as the particle remains below the step, but no lag when descending the step. For smaller step heights, the particle easily jumps up and hence does not significantly lag the trap, whereas for larger step heights the equilibrium position distribution never achieves significant probability above the step, and hence the nonequilibrium distribution never lags the equilibrium one.
Such an intermediate step height is found in the $\sim11\kT$ conformational change driving the overwhelmingly forward stepping of the biomolecular transport motor kinesin~\cite{hackney05}.

We also found that protocols must end an intermediate distance from the potential step for time asymmetry to exceed linear response. Protocols ending far from the step had higher dissipation, leaving less difference between the linear response time asymmetry and the maximum possible time asymmetry. 
Protocols ending close to the step never move the equilibrium distribution above the step, 
precluding the possibility of distinct work distributions.
Our theory predicts that the range $\sqrt{8\Delta E/k} < \Delta x \ltsim \exp(\Delta E)/(D k u)$ corresponds to time asymmetry exceeding linear response. 
This preference for an intermediate initial distance from the step is suggestively consistent with experiments demonstrating the impact of kinesin neck-linker length on its transport anisotropy~\cite{isojima16}: compared to wild type, artificially shortened or lengthened neck-linkers each produce a lower ratio of forward to backward steps. 

The system remaining below the step while the protocol completes suggests an intuitive perspective:
exceeding linear response can result from a separation of timescales during the forward protocol but not the reverse. The protocol period, defining the timescale for change in the equilibrium states, must be significantly shorter than the timescale for the particle to jump up the step (the equilibration timescale for the forward protocol), but significantly longer than the timescale for the particle jumping down the step.

During the operation of kinesin, the rear head unbinding followed by rebinding at a forward site is thought to be a largely irreversible transition~\cite{clancy11}. Neck-linker docking, a conformational change leading to a forward step of the unbound kinesin head, is much faster than the timescale of head diffusion and binding of the forward microtubule site~\cite{muretta15b,isojima16}. 
A `reverse protocol' with sufficient assisting force could eliminate or potentially reverse this separation of timescales, satisfying our posited criteria for efficient dissipation.

\begin{acknowledgments}
This work was supported by a Natural Sciences and Engineering Research Council of Canada (NSERC) Discovery Grant (DAS) and by funds provided by the Faculty of Science, Simon Fraser University through the President's Research Start-up Grant (DAS), and was enabled in part by support provided by WestGrid (www.westgrid.ca) and Compute Canada Calcul Canada (www.computecanada.ca). The authors thank John Bechhoefer and Nancy Forde (SFU Physics), Leonid Chindelevitch (SFU Computing Science), and Bingyun Sun (SFU Chemistry) for useful discussions and feedback.
\end{acknowledgments}

\appendix

\section{Non-dimensionalization}
\label{sec:nondim}

The energy $\Edim$ of a particle at position $\xdim$ in quadratic trap with spring constant $\kdim$ centered at $\xdim = \xdim_{\textrm{trap}}$ with a step potential of height $\DeltaEdim$ is
\begin{equation}
\Edim = \frac{1}{2}\kdim\left(\xdim - \xdim_{\textrm{trap}}\right)^2 + \DeltaEdim\Theta(\xdim).
\end{equation}
We non-dimensionalize all quantities using energy scale $H$ and length scale $\ell$:
\begin{equation}
\frac{\Edim}{\Escale} = \frac{1}{2}\frac{\kdim\lscale^2}{\Escale}\frac{\left(\xdim - \xdim_{\textrm{trap}}\right)^2}{\lscale^2} + \frac{\DeltaEdim}{\Escale}\Theta\left(\frac{\xdim}{\lscale}\right).
\end{equation}
We choose $\Escale=\kT$, where $k_{\rm B}$ is Boltzmann's constant and $T$ is absolute temperature, so that all energies are in units of $\kT$. $\lscale$ is any length scale, e.g. nm or $\mu$m, or a system lengthscale such as particle size. We set $E \equiv \Edim/\Escale$, $k \equiv \kdim\lscale^2/\Escale$, $x \equiv \xdim/\lscale$, and $\DeltaEdim \equiv \Delta E/\Escale$ to give the energy equation in the main text. We also non-dimensionalize time using a timescale $\tscale$: $t = \tdim/\tscale$.

\section{Numerical simulation details}
\label{sec:numdet}

The forward protocol begins with the particle in equilibrium at trap minimum $\xT = -\Delta x/2$ below the step. The trap minimum moves a distance $\Delta x$ to $\xT = \Delta x/2$ over a time period $\Delta t$. The trap moves at a constant velocity, such that $\xT^F(t) = -\Delta x/2 + \Delta x\cdot t/\Delta t$. 
Reverse protocols begin at equilibrium with $\xT = \Delta x/2$ and proceed with negative trap velocity such that $\xT^R(t) = \Delta x/2 - \Delta x\cdot t/\Delta t$. 

Particle and trap position are evolved using a Gillespie algorithm~\cite{gillespie77,gillespie07}. Particle states are discretized with $\delta x_p = 0.05$. The particle takes steps either left or right, with rates described by the Metropolis criterion. The rate is $\Gamma = \Gamma_0\exp(-\delta E)$ for energy difference $\delta E>0$ between current and proposed states, flat-landscape transition rate $\Gamma_0 = 2D/(\delta x_p)^2$, and diffusion coefficient $D = 1/2$. The rate is $\Gamma = \Gamma_0$ for $\delta E \leq 0$.
The trap translates in steps of $\delta \xT = 10^{-4}$, to maintain a velocity $u=\delta \xT/\delta t$, with $1/\delta t$ defining the rate for the Gillespie algorithm. The small $\delta \xT$ leads to high rates $1/\delta t = u/\delta \xT$ in the Gillespie algorithm, such that the trap moves nearly deterministically.

The forward protocol free energy change is $\Delta F = F[\xT^{F}(\Delta t)] - F[\xT^{F}(0)]$, for equilibrium free energy $F[\xT] \equiv -\ln \sum_x \exp\left[-E(x,\xT)\right]$.

\section{Work distributions}
\label{sec:workdist}

\subsection{Forward work distributions}

For the forward process, the distribution of times for a particle to jump up and remain above the step is
\begin{equation}
\label{eq:jumpup}
P_{\textrm{jump up}}(t) = P_{\textrm{down step}}(t)\Gamma_{\textrm{down}\to  \textrm{up}}(t) \ ,
\end{equation}
where
\begin{equation}
\label{eq:downstep}
\frac{\md P_{\textrm{down step}}(t)}{\md t} = -\Gamma_{\textrm{down}\to \textrm{up}}(t) P_{\textrm{down step}}(t),
\end{equation}
with initial condition $P_{\textrm{down step}}(t=\Delta t/2)=1$. 
We split the rate $\Gamma_{\textrm{down}\to \textrm{up}}(t) = \Gamma_{\textrm{jump}}(t) P_{\textrm{stay}}(t)$ at which a particle permanently jumps up the step into a product of two terms: $\Gamma_{\textrm{jump}}(t)$ is the rate at which a particle arrives at the step and jumps up, and $P_{\textrm{stay}}(t)$ is the probability that a particle, once it has jumped up the step, will remain above the step until the end of the protocol. 

To find $\Gamma_{\textrm{jump}}(t)$, we consider the timescale of attempted jumps up,  $\tau_{\textrm{attempt}}(t) = \langle x^2\rangle_t/(2D)$. A fraction $\exp(-\Delta E)$ of the attempts succeed so that the rate of jumping up the step is $\Gamma_{\textrm{jump}}(t) = \exp(-\Delta E)/\tau_{\textrm{attempt}}(t)$. $\langle x^2\rangle_t$ is the average squared distance of the particle from the step while it is stuck on the low side, with the trap minimum above the step,
\begin{equation}
\langle x^2\rangle_t = \frac{1}{Z(t)}\int_0^{\infty}x^2\exp\left\{-\half k\left[x-\xT(t)\right]^2\right\}dx,
\end{equation}
where
\begin{equation}
\label{eq:partition}
Z(t) \equiv \sqrt{\frac{\pi}{2k}}\left\{1 + \textrm{erf}\left[\sqrt{\frac{k}{2}}\xT(t)\right]\right\}.
\end{equation}
Integrating gives
\begin{equation}
\label{eq:avgxsquared}
\langle x^2\rangle_t = \xT^2(t) + \frac{1}{k} + \frac{\sqrt{\frac{2}{k \pi}}\xT(t)\exp\left[-\half k\xT^2(t)\right]}{1 + \textrm{erf}\left[\sqrt{\frac{k}{2}}\xT(t)\right]}.
\end{equation}

To find $P_{\textrm{stay}}$(t),
we model escape from the trap, when centered above the step, as escape from a truncated quadratic trap with potential
\begin{equation}
V(x) \equiv
\begin{cases}
\half k(x-\xT)^2, & \textrm{if } x\leq0\\
-\infty, & \textrm{if } x>0.
\end{cases}
\end{equation}
The rate of escape from this trap is~\cite{hummer03}
\begin{equation}
\label{eq:f4}
\Gamma_{\textrm{escape}}(t) = \frac{Dk^{3/2}}{\sqrt{2\pi}}|\xT(t)|\exp\left[-\half k\xT^2(t)\right].
\end{equation}
For a particle that has just jumped into the trap at time $t_{\textrm{jump}}$, the probability that it remains in the trap is
\begin{equation}
\label{eq:f3}
\frac{\md P_{\textrm{in trap}}(t)}{\md t} = -\Gamma_{\textrm{escape}}(t) P_{\textrm{in trap}}(t),
\end{equation}
with $P_{\textrm{in trap}}(t_{\textrm{jump}}) = 1$ (if the particle jumped into the trap at time $t_{\textrm{jump}}$, the probability that the particle is in the trap is unity at this time). 
Solving for the probability that the particle will remain in the trap until the end of the protocol at time $\Delta x/u$, given that it jumped up the step at time $t_{\textrm{jump}}$,
\begin{subequations}
\begin{align}
P_{\textrm{stay}}&(t_{\textrm{jump}}) = P_{\textrm{in trap}}(\Delta x/u) \\
\label{eq:stay} &= \exp\left\{-\frac{D\sqrt{k}}{u\sqrt{2\pi}}\left[e^{-\half k\xT^2(t_{\textrm{jump}})} - e^{- \half k(\Delta x/2)^2}\right]\right\}.
\end{align}
\end{subequations}
Combining $\Gamma_{\textrm{jump}}(t)$ (in-line in paragraph following Eq.~\ref{eq:downstep}) and $P_{\textrm{stay}}(t)$ (Eq.~\ref{eq:stay}) gives
\begin{align}
\label{eq:huge}
\Gamma_{\textrm{down}\to \textrm{up}}(t) &= 2De^{-\Delta E}\times \\
\exp &\left[-\frac{D\sqrt{k}}{u\sqrt{2\pi}}\left(e^{-\half k \xT^2(t)} - e^{-\half k(\Delta x/2)^2}\right)\right] \bigg/ \nonumber\\
&\left\{\xT^2(t) + \frac{1}{k} + \frac{\sqrt{\frac{2}{k\pi}}\xT(t) e^{-\half k\xT^2(t)}}{1 - \textrm{erf}\left[\sqrt{\frac{k}{2}}\xT(t)\right]}\right\} \ .\nonumber
\end{align}

When the forward protocol finishes at $\xT = \Delta x/2$, the particle remains below the step with probability $P_{\textrm{down step}}(\Delta t)$. Such trajectories form the distinct peak at high work values in the forward work distribution.

To get work distributions from these distributions of the time the particle jumps up the step, 
we need 
the forward protocol work accumulated to time $t$,
\begin{equation}
\label{eq:workforward}
W_{\textrm{F}}(t) = \int_{0}^{t}k\langle x - x_{\textrm{trap}}^{\textrm{F}}(t')\rangle \frac{\md x}{\md t'}\; \md t'.
\end{equation}
The average deviation 
of the particle from the trap minimum is
\begin{subequations}
\begin{align}
\langle x - \xT(t) \rangle_t &= \frac{1}{Z(t)}\int_0^{\infty}x\exp\left\{-\half k\left[x - \xT(t)\right]^2\right\}\md x \\
&= \frac{\sqrt{\frac{2}{k\pi}}\exp\left[-\half k\xT^2(t)\right]}{1 - \textrm{erf}\left[\sqrt{\frac{k}{2}}\xT(t)\right]} \ .
\end{align}
\end{subequations}
where $Z(t)$ is defined in Eq.~\ref{eq:partition}.
The work distributions derived above are shifted and convolved with a Gaussian distribution according to linear response theory~\cite{mazonka99} (details below) to produce the semi-analytical distributions in Fig.~2.

\subsection{Reverse work distributions}

For the reverse process, the distribution of times $t$ (after the beginning of the protocol) for a particle to jump down the step is
\begin{equation}
\label{eq:jumpdown}
P_{\textrm{jump down}}(t) = P_{\textrm{up step}}(t)\Gamma_{\textrm{up}\to \textrm{down}}(t) \ ,
\end{equation}
where
\begin{equation}
\label{eq:upstep2}
\frac{\md P_{\textrm{up step}}(t)}{\md t} = -\Gamma_{\textrm{up}\to \textrm{down}}(t) P_{\textrm{up step}}(t) \ .
\end{equation}
$\Gamma_{\textrm{up}\to \textrm{down}}(t)$ is the escape rate from Eq.~\ref{eq:f4}. Solving Eq.~\ref{eq:upstep2} with $P_{\textrm{up step}}(0) = 1-P_0$ gives
\begin{equation}
\label{eq:upstep}
\begin{array}{l}
\frac{P_{\textrm{up step}}(t)}{1-P_0} = \exp\left\{-\frac{D}{u}\sqrt{\frac{k}{2\pi}}
e^{-k(\Delta x)^2/8}\left[e^{-kut\left(ut - \Delta x\right)/2} - 1\right]\right\}.
\end{array}
\end{equation}
$P_0= 1/(1+Q)$ is the probability that the system is below the step in equilibrium at the beginning of the reverse protocol, with 
\begin{equation}
\label{eq:q}
Q \equiv e^{-\Delta E}\frac{1 + \textrm{erf}\left(\sqrt{\frac{k}{2}}\frac{\Delta x}{2}\right)}{1 - \textrm{erf}\left(\sqrt{\frac{k}{2}}\frac{\Delta x}{2}\right)} \ .
\end{equation}
$P_0$ quantifies the size of the distinct high-work peak of the reverse work distribution.

The reverse work is
\begin{equation}
\label{eq:workreverse}
W_{\textrm{R}}(t) = \int_{0}^{t}k\langle x - \xT^{\textrm{R}}(t')\rangle_{t'} \frac{\md x}{\md t'}\; \md t' \ .
\end{equation}

\subsection{Convolving and shifting work distributions}
Eq.~\ref{eq:jumpup} gives a distribution of times to jump up the step for the forward protocol, and Eq.~\ref{eq:jumpdown} the distribution of times to jump down the step for the reverse protocol. 
We also consider the probabilities that a particle does not jump up the step by the end of the forward protocol (see discussion after Eq.~\ref{eq:huge}) and that a particle starts the protocol in equilibrium down the step for the reverse protocol (see in-line equation after Eq.~\ref{eq:upstep}). Together, these form forward and reverse time distributions for jumping up and down the step. Eqs.~\ref{eq:workforward} and~\ref{eq:workreverse} transform these time distributions into work distributions. 

To complete our description of the work statistics, we also account for
the work exerted while the particle remains on one side of the step.
We convolve the low-work peaks with a Gaussian of variance
\begin{equation}
\sigma^2 = \frac{2(\Delta x)^2}{D\Delta t}\left[1 + \frac{1}{Dk\Delta t}\left(e^{-Dk\Delta t} - 1\right)\right] \ ,
\end{equation}
and mean $\sigma^2/2$, 
that exactly describes the work fluctuations 
for a quadratic trap translating at constant velocity on a flat energy landscape~\cite{mazonka99}.
The high-work peaks, which correspond to trajectories where the particle remains below the step for the entire protocol,
are convolved and shifted using half the above variance, as such a particle will travel roughly $\Delta x/2$, half the distance $\Delta x$ travelled by a particle that jumps up the step with sufficient time to equilibrate before the protocol concludes.

\section{Criteria for predicting on, below, or above linear response}
\label{sec:criteria}

First we derive the condition that the forward work distribution is stretched out. We expect this when the timescale of a particle jumping up the step,
\begin{equation}
\label{eq:timescale1}
\Gamma_{\textrm{jump}}^{-1}(t) = \frac{\langle x^2\rangle_t}{2D\exp\left(-\Delta E\right)} \ ,
\end{equation}
is longer than the timescale $\tC - (\Delta t/2)$ to remain up the step once it jumps, 
which satisfies $P_{\textrm{stay}}(\tC) = 0.5$ for $P_{\textrm{stay}}(t)$ defined in Eq.~\ref{eq:stay}.

To find the timescale $\Gamma_{\textrm{jump}}^{-1}(t)$ in Eq.~\ref{eq:timescale1}, we start with $\langle x^2\rangle_t$, calculated in Eq.~\ref{eq:avgxsquared}. Assuming the trap is far from the step compared to the trap width, i.e. 
$\half k\xT^2(t) \gg 1$ (necessary to avoid immediate escape of the particle), we find $\langle x^2\rangle_t \simeq 3/[k^2\xT^2(t)]$. Using $\xT(t) = u(t - \Delta t/2)$ and inserting into Eq.~\ref{eq:timescale1} produces
\begin{equation}
\label{eq:tscale1}
\Gamma_{\textrm{jump}}^{-1}(t) = \frac{3}{2D(ku)^2 (t - \Delta t/2)^2\exp(-\Delta E)}.
\end{equation}

To determine $\tC - (\Delta t/2)$, we assume the particle jumps  when the trap minimum is far from the end of the protocol, i.e.
$ut \ll \Delta x/2$,
and solve Eq.~\ref{eq:stay} for $t-\Delta t/2$ to give
\begin{equation}
\label{eq:tscale2}
t - \frac{\Delta t}{2}= \sqrt{-\frac{2}{k u^2}\ln\left[-\frac{u}{D}\sqrt{\frac{2\pi}{k}}\ln P_{\textrm{stay}}(t)\right]} \ .
\end{equation}
Imposing $t=\tC$ and $P_{\textrm{stay}}(\tC)=\half$, and substituting
into $\Gamma_{\textrm{jump}}^{-1}(t) \gtrsim \tC - (\Delta t/2)$ gives (after rearrangement)
\begin{equation}
\exp\left(2\Delta E\right) \gtrsim \frac{32}{9}\frac{D^2 k}{u^2} \ln \left( \frac{D}{u\ln 2} \sqrt{\frac{k}{2\pi}} \right) \ .
\end{equation}
Assuming the logarithmic term is order unity produces
\begin{equation}
\Delta E \gtrsim \half\ln\left(\frac{32}{9}\frac{D^2 k}{u^2}\right) \ .
\end{equation}

Next, we derive the condition
for the forward work distribution to have a significant high-work peak.
We expect this when the timescale for jumping up the step is longer than the remaining protocol time (half the total protocol time elapses after the trap minimum passes the step),
\begin{equation}
\Gamma_{\textrm{jump}}^{-1}(t) \gtsim \frac{\Delta t}{2} \ .
\end{equation}
Substituting $\Gamma_{\textrm{jump}}$ from Eq.~\ref{eq:tscale1} and $\tC-\Delta t/2$ from Eq.~\ref{eq:tscale2}
and rearranging:
\begin{equation}
\exp(\Delta E) \gtsim \frac{2}{3} D k\Delta t\ln\left( \frac{D}{u\ln 2} \sqrt{\frac{ k}{2\pi}} \right) \ .
\end{equation}
Approximating
the logarithmic term 
as unity gives
\begin{equation}
\Delta E \gtrsim \ln D k\Delta t \ .
\end{equation}

Next, we derive the condition
for the reverse work distribution to have a significant high-work peak, which occurs when the initial equilibrium distribution has significant support down the step.
At the beginning of the reverse protocol, $\xT = \Delta x/2$. The energy at the trap minimum is $\Delta E$, and the energy immediately below the step is $\half k(\Delta x/2)^2$. Thus the equilibrium probability density immediately below the step exceeds the probability density at the trap minimum when
\begin{equation}
\Delta E > \half k\left(\frac{\Delta x}{2}\right)^2.
\end{equation}

Next, we derive the condition 
for the reverse work distribution to have a dominant peak at high work values. At the beginning of the reverse protocol, we require that the equilibrium probability immediately below the step is much higher (arbitrarily set to $100\times$) than the probability at the trap minimum:
\begin{equation}
\exp\left[-\half k\left(\frac{\Delta x}{2}\right)^2\right] > 100\exp(-\Delta E),
\end{equation}
which rearranges to
\begin{equation}
\Delta E > \half k\left(\frac{\Delta x}{2}\right)^2 + \ln 100\ .
\end{equation}

Finally, we define the forward and reverse distributions as effectively distinct when they have only 0.1\% overlap, i.e. when 
$x_{\textrm{F,edge}} > x_{\textrm{R,edge}}$ for 
\begin{subequations}
\begin{align}
0.999 &= \int_{x_{\textrm{F,edge}}}^{\infty}P_F(W)dW \\
0.999 &= \int_{-\infty}^{x_{\textrm{R,edge}}}P_R(W)dW \ .
\end{align}
\end{subequations}

\section{Maximizing time asymmetry}
\label{sec:maximizing}

The forward and reverse work distributions can each develop two peaks, resulting from two different classes of trajectories. The low-work peak reflects trajectories where the particle stays near the trap as it crosses the step. The high-work peak results from the particle remaining below the step for the entire protocol.

For small step heights $\Delta E$, both the forward and reverse work distributions only have a low-work peak. As the step height increases, the forward work distribution develops a high-work peak, and as the step height increases further, the reverse work distribution also develops such a second peak. The time asymmetry $A$ increases as the trajectory distributions for the forward and reverse protocols become more distinct, therefore increases for higher probabilities of the forward high-work peak, and subsequently reverses for higher probabilities of the reverse high-work peak.

The weight of the forward high-work peak is the probability that a particle remains below the step at the end of the forward protocol. The rate at which a particle jumps up the step is
\begin{equation}
\Gamma_{\textrm{down}\to \textrm{up}}(t) = \frac{2D\exp(-\Delta E)}{\langle x^2\rangle_t}P_{\textrm{stay}}(t) \ .
\end{equation}

We assume that $P_{\textrm{stay}}(t) \simeq 1$ for times $t>\tC$, i.e. $P_{\textrm{stay}}(t) \simeq \Theta(t-\tC)$, with the time $\tC$ defined by $P_{\textrm{stay}}(\tC) = \half$ from Eq.~\ref{eq:tscale2}. 
Using $\xT(t) = u(t-\Delta t/2)$ and the inline equation following Eq.~\ref{eq:timescale1}, $\langle x^2\rangle_t = 3/[k^2\xT^2(t)]$, gives
\begin{equation}
\label{eq:simplifiedtransition}
\Gamma_{\textrm{down}\to \textrm{up}}(t) \simeq \frac{2}{3}D(k u)^2 (t-\Delta t)^2\exp(- \Delta E) \ .
\end{equation}
We substitute Eq.~\ref{eq:simplifiedtransition} into Eq.~\ref{eq:downstep} and integrate:
\begin{align}
\int&_{P_{\textrm{down step}}(\tC)}^{P_{\textrm{down step}}(t)}\frac{\md P'_{\textrm{down step}}}{P'_{\textrm{down step}}} =\\
&-\frac{2}{3}D(ku)^2\exp(-\Delta E)\int_{\tC}^{t}(t' - \Delta t/2)^2 \md t' \ , \nonumber
\end{align}
which for $P_{\textrm{down step}}(\tC)\simeq 1$ gives the probability of a particle remaining below the step for the entire protocol:
\begin{align}
P&_{\textrm{down step}}(\Delta t) \simeq\\
&\exp\left[-\frac{2}{9}D(ku)^2e^{-\Delta E}\left[(\Delta t/2)^3 - (\tC - \Delta t/2)^3\right]\right] \ . \nonumber
\end{align}

The weight of the reverse low-work peak is $P_0=1/(1+Q)$,
for $Q$ defined in Eq.~\ref{eq:q}.

Only at intermediate dissipation $h$ is there any room to $\ARoom(h)$ to improve upon the time asymmetry $\ALR(h)$ of linear response, before hitting the maximum possible time asymmetry, $\ALimit(h)$. 
Here, by examining the forward high-work peak and the reverse low-work peak, we crudely estimate the dissipation and thereby
determine this difference $\ARoom(h) = \ALimit(h) - \ALR(h)$.

We estimate the mode of the forward high-work peak as $W = \half k(\Delta x/2)^2$, the energy difference between the particle stuck behind the step at the end of the forward protocol when $\xT = \Delta x/2$, and the particle at the step when the trap minimum crosses the step, which dominates the work accrued with the trap below the step. 

To estimate the mode of the reverse low-work peak, we find the trap position at which the particle is most likely to jump down the step during the reverse protocol, and calculate the subsequent work done as the trap minimum approaches and crosses the step, which dominates the work accumulated while the particle and trap minimum are on the same side of the step. 
Assuming $P_{\textrm{up step}} \simeq 1$, the probability of jumping down the step (Eq.~\ref{eq:jumpdown}) simplifies to
\begin{equation}
P_{\textrm{jump down}}(\xT) \simeq \frac{Dk^{3/2}}{\sqrt{2\pi}}\xT\exp\left(-\half k\xT^2\right) \ .
\end{equation}
Maximizing $P_{\textrm{jump down}}(\xT)$ by setting $\md P_{\textrm{jump down}}(\xT)/\md \xT = 0$ gives $\xT^{\textrm{peak}} = k^{-1/2}$. The corresponding work while the particle and trap are on opposite sides of the step is $\half k (\xT^{\textrm{peak}})^2$. 

The dissipation is half the difference between the two peaks,
\begin{subequations}
\begin{align}
h(k,\Delta x) &= \half\left[\half k\left(\frac{\Delta x}{2}\right)^2 - \half k(\xT^{\textrm{peak}})^2\right] \\
&= \frac{1}{4}\left[k \left(\frac{\Delta x}{2}\right)^2 - 1\right] \ .
\end{align}
\end{subequations}
We use this dissipation estimate to determine the room above linear response $\ARoom[h(k,\Delta x)]$.
The color map of Fig.~3e plots $(1-P_0)P_{\textrm{down step}}(\Delta t)\ARoom$.

\bibliography{DissipationIrreversibility}

\end{document}